\documentclass[12pt, a4paper]{article}

\usepackage[T1]{fontenc}
\usepackage{times}
\usepackage{amsmath, amssymb, amsfonts}   

\usepackage{graphicx}
\usepackage{caption, subcaption}

\usepackage[margin=2cm]{geometry}
\usepackage[onehalfspacing]{setspace}

\usepackage[round, sort & compress]{natbib}
\usepackage{hyperref}

\title{Detection of Alzheimers Disease from MRI using Convolutional Neural Networks, Exploring Transfer Learning And BellCNN}	
\author{GuruRaj Awate \\ gururaj.formal@gmail.com}

\begin{document}
	\maketitle
		\begin{abstract}
			
			There is a need for automatic diagnosis of certain diseases from medical images that could help medical practitioners for further assessment towards treating the illness. Alzheimers disease is a good example of a disease that is often misdiagnosed. Alzheimers disease (Hear after referred to as AD), is caused by atrophy of certain brain regions and by brain cell death and is the leading cause of dementia and memory loss [1]. MRI scans reveal this information but atrophied regions are different for different individuals which makes the diagnosis a bit more trickier and often gets misdiagnosed [1, 13]. We believe that our approach to this particular problem would improve the assessment quality by pre-flagging the images which are more likely to have AD. We propose two solutions to this; one with transfer learning [9] and other by BellCNN [14], a custom made Convolutional Neural Network (Hear after referred to as CNN). Advantages and disadvantages of each approach will also be discussed in their respective sections. 
			
			The dataset used for this project is provided by Open Access Series of Imaging Studies (Hear after referred to as OASIS) [2, 3, 4], which contains over 400 subjects, 100 of whom have mild to severe dementia. The dataset has labeled these subjects by two standards of diagnosis; Mini-Mental State Examination (Hear after referred to as MMSE) and Clinical Dementia Rating (Hear after referred to as CDR). These are some of the general tools and concepts which are prerequisites to our solution; CNN [5, 6], Neural Networks [10] (Hear after referred to as NN), Anaconda bundle for python, Regression, Tensorflow [7].
	\end{abstract}

	\section{INTRODUCTION}
		AD is a neurological disorder that causes memory loss and dementia and It is mainly observed in elderly individuals over the age of 60 but can also be caused by concussions or traumatic brain injuries [1, 2, 3, 4, 13]. It causes brain cells to die and spreads the damage across the brain, in some severe cases rendering an individual unable to perform daily tasks necessary for ones survival. It is also considered as a neurodegenerative type of dementia [1, 13]. There are a few ways to diagnose and detect the disease manually, such as examination of MRI scans and MMSE scores; these manual tests can also be expressed in terms of CDR [2, 3]. But identifying distinctions between AD brain and a normal functioning brain in elderly individuals over the age of 75 is difficult because they share similar brain patterns and image intensities [13].
	
		\begin{figure}[!htbp]
			\centering
			\includegraphics{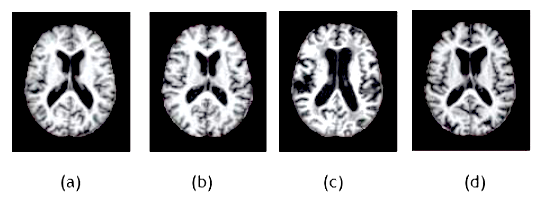}
			\caption{Shows comparison of Normal and Demented MRIs [2]}
			(a) no dementia,
			(b) very mild dementia,
			(c) mild dementia,
			(d) moderate dementia.
		\end{figure}
	
	\section{DATA ACQUISITION}
		The dataset being used is obtained from OASIS which openly provides the relevant data containing cross-sectional MRI scans for demented and non-demented individuals ranging from the age of 18 years to 96 years. The dataset consists of 416 subjects, all of which are right handed. The dataset comprises of standardized T1-weighted Images; from which 100 subjects have mild to moderate dementia and 198 subjects which are all over the age of 60.
		
		\begin{table}[!htbp]
				\centering
				\caption{CDR AND DEMENTIA [2, 3]}
				\label{tab:2.1}
				\begin{tabular}{p{1cm}p{1cm}p{4cm}}
						\hline 
						Sr. & CDR & Dementia Status \\ 
						\hline
						1.  & 0   & No dementia \\
						2. & 0.5  & Very mild dementia \\
						3.  & 1   & Mild dementia \\
						4.  & 2   & Moderate dementia \\
						5.  & 3   & Severe dementia\\
						\hline
				\end{tabular} 	
		\end{table}  
	
		To avoid overfitting the dataset also includes 218 non-demented subjects from the age group of 18 to 59 years. The dataset also has 98 elderly subjects over the age of 60 with no dementia and an CDR of 0. For demented class of individuals, CDR scores are as shown in table.
		
		\begin{table}[!htbp]
			\centering
			\caption{DEMOGRAPHICS OF DEMENTIA IN OASIS DATASET [2, 3]}
			\label{tab:2.2}
			\begin{tabular}{p{1cm}p{2cm}p{1cm}p{2cm}p{2cm}p{2cm}}
				\hline 
				Sr.   &    Age &      N     &    CDR = 0.5  &   CDR = 1    &   CDR = 2\\
				\hline 
				1.   &     60-69 &    15  &      12 &           3  &           -\\
				2.    &    70-79  &   48   &     32      &      15     &       1\\
				3.     &   80-89   &  32   &     22      &      9         &    1\\
				4.   &     90-96 &    5     &    4          &   1       &      -\\
				\hline 
				Total & &  100 &  70  &   28 &  2\\
			\end{tabular} 	
		\end{table} 
	
		We will use processed images provided within the dataset to train our model but some pre-labeling is also required; labeling done on the basis of CDR = 0 indicating no dementia and CDR > 0 indicating dementia. Some of the data is preserved as a test dataset. The additional data presented in the accompanying CSV file will be used to gain some insight and metadata about the dataset which would support the data wrangling process.
	
	\section{TECHNOLOGIES}
	
			\subsection{Convolutional Neural Network:}
			CNNs are mainly used for graphical data in recognition, processing of images. These are inspired by the human visual systems and were created by Yann LeCun [5, 6]. The input image is broken down into smaller chunks known as local receptive fields. CNN comes with 3 main layers each having a specific function to perform namely, convolutional layer, pooling layer and fully connected layer. Convolutional layer (Here after referred to as CL) has set of learnable filters that can be applied on input images. It also consists of 3 dimensions and spatial portions [8].
			
			Each operation requires 4 hyper parameters:\\
			K, number of filters; Vector of weights to be applied to input image,\\
			F, their spatial extent; 3 Dimensions- width, height and depth.\\
			S, the size of stride; Amount of displacement for the filter, in pixels.\\
			P, the amount of (zero) padding; It control dis-proportionate input to fit the filter.
			
			Application of filter or a convolution operation involves Shift, Scale, Rotate, Other transformations. To find patterns, CNN takes input images as $W_1 * H_1 *D_1.$ \\
			
			Then it outputs Next image as:$\\
					W_2 = (W_1 - F) /S + 1\\
					H_2 = (H_1 - F) /S + 1\\
					D_2 = D_1
				$
			
			While CL handles image filtration, Pooling layer (Here after referred to as PL) deals with reducing the sample size and controlling overfitting. It is usually placed between CL and Fully connected layer (Here after referred to as FCL); which is a classic neural network with every node interconnected with every other node [5, 6, 13]. Rectified Linear unit (Here after referred to as RELU) function is used to obtain the unsigned segments in the input space. In CNNs we will use them as our activation function. RELUs also have a distinguishable advantage; that it does not require the input to be normalized which stops them from becoming saturated.
			
			\subsection{Tensorflow:}
			Tensorflow is an open Source Deep Learning API, it was Developed by Googles Google-Brain Team. It Uses 3-D Data-store instead of a Vector, Known as Tensor, Tensor stores more amount of data in one go. Tensorflow require users to create a Network graph, and the tensor variables flow in that graph, hence the name Tensorflow, Tensorflow also provides pre-trained CNN known to everyone as Inception [7, 8, 9], which makes transfer learning possible for people who do not have GPUs available with them for High performance computing (Here after referred to as HPC). Tensorflow is highly scalable and can be ported to other platforms for inference and deployment; as internally it runs on low level programming languages such as C++. Tensorflow Model once trained, doesnt require HPC as the learned features are preserved. 
			
			Tensorflow does this by using protocol buffers, often referred as protobuffs with a (.pb) extension. Essentially, they are a way to store information about graphs in the form of network structure and weights; we can easily obtain protobuffs by freezing a working tensorflow graph [7]. Creating visualizations in tensorflow is done through tensorboard. It is a utility used with training summaries to create visualizations like graphs, histograms and other similar techniques of visual representation. Training summaries are stored in a file while training the model, they contain information such as training step, accuracy, loss and other useful parameters [7].
			
			\section{PROPOSED SYSTEMS}
			
			\subsection{Transfer Learning: }
			Transfer Learning (Here after referred to as TL) is a process of utilizing a ML model which is trained for different purpose on different data. TL does this by retraining the frozen model on the data you wish to work on. The thing to notice here is that only the FL and The final layer is optimized for the new data, not the entire model. So its highly likely that the model on some occasions will miss and give out an incorrect result [8,9].
			
			In our scenario, we will use Tensorflow for poets; which allows us to retrain and repurpose two different kinds of models namely Inception and MobileNet; with an exact same procedure [9]. While retraining it computes bottlenecks for every image; Bottlenecking in Tensorflow is a technique similar to memoization which helps reduce redundant computations. The FL is optimized on these new images and final regression layer is adjusted according to the specification; here in our case, It consists of a binary outcome. These retrained graphs can be frozen as a protobuff (.pb) file and reused with desired deployment strategy. 
			
			\begin{figure}[!htbp]
				\centering
				\includegraphics[width=\textwidth,height=\textheight,keepaspectratio]{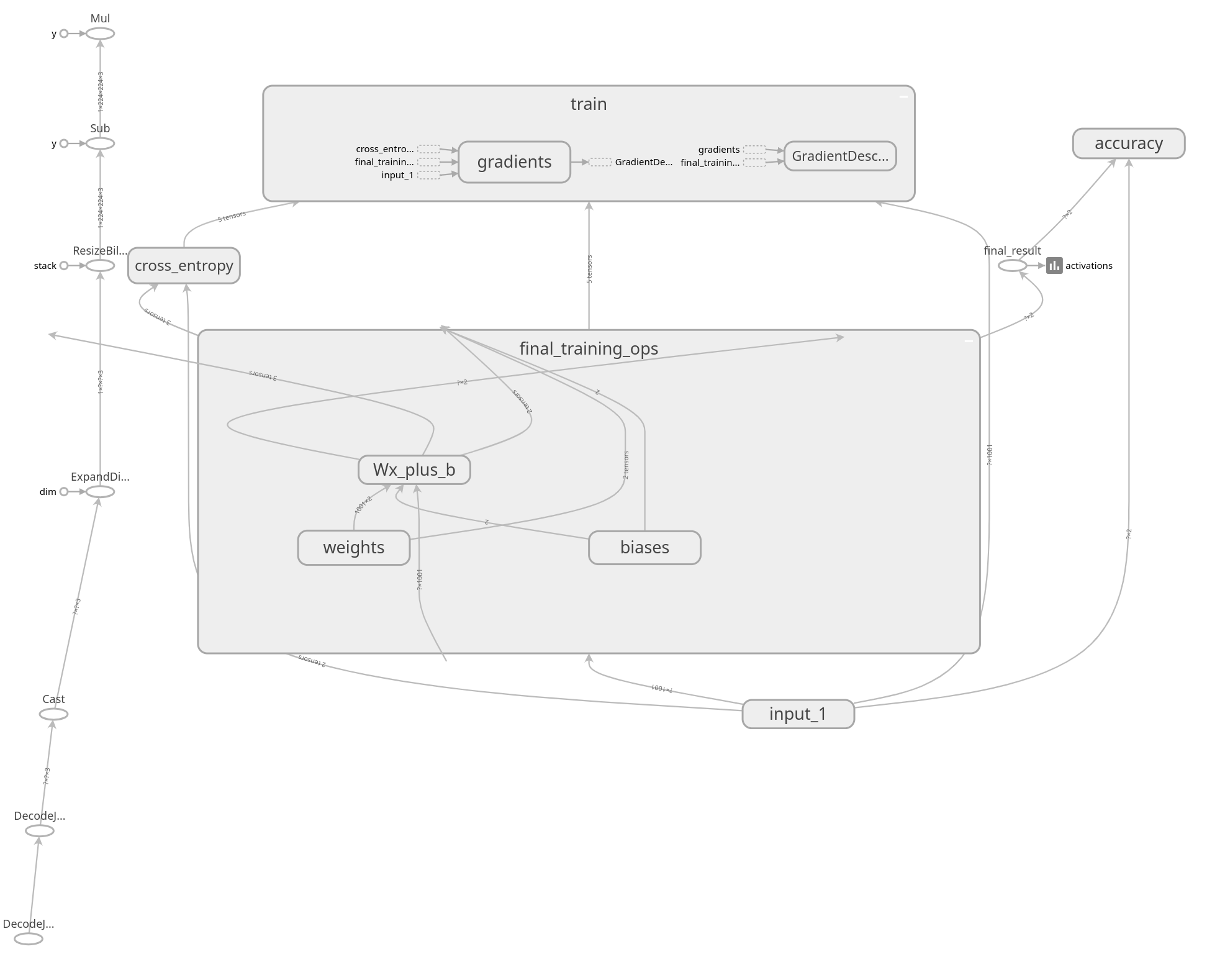}
				\caption{Final operations performed while Retraining}
				This graph is a Tensorboard visualization of final operations performed while retraining; This operation is same for both, MobileNet and Inception.
			\end{figure}
			
			\begin{figure}[!htbp]
				\centering
				\includegraphics[width=\textwidth,height=\textheight,keepaspectratio]{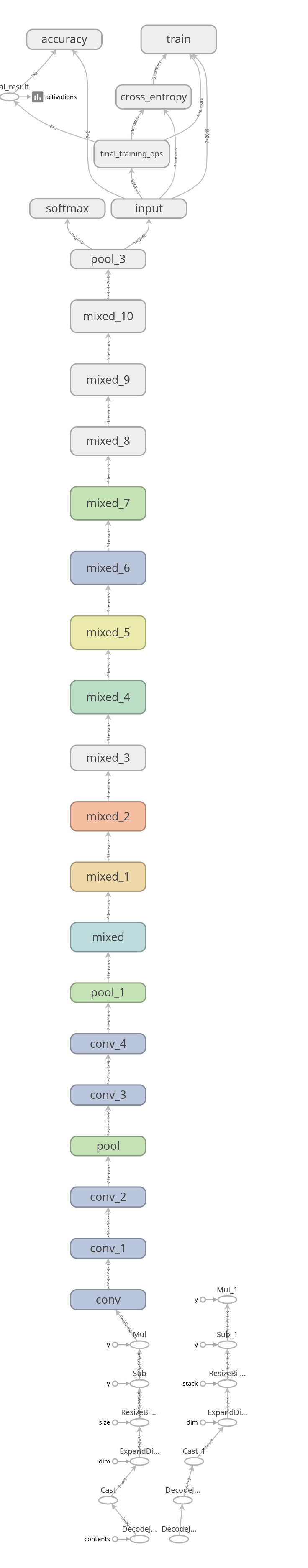}
				\caption{The Tensorboard visualization for our frozen graph of retrained Inception model.}
			\end{figure}
		
		\begin{figure}[!htbp]
			\centering
			\includegraphics[width=\textwidth,height=\textheight,keepaspectratio]{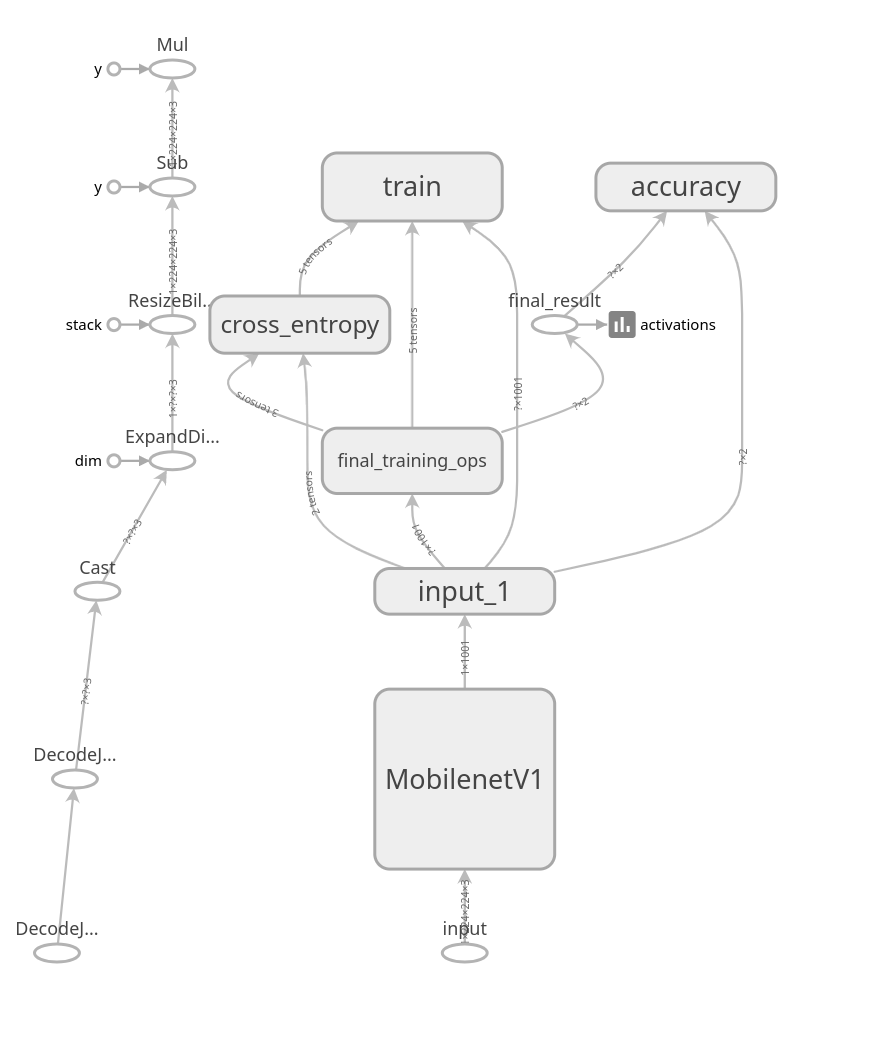}
			\caption{The Tensorboard visualization for our frozen graph of retrained MobileNet model.}
		\end{figure}
			\pagebreak
		    Retraining on Inception:\\
		    	For Inception the accuracy was close to 86\%.	\\
			$
			INFO:tensorflow:Final\ test\ accuracy = 85.7\%\ (N=98).\\
			INFO:tensorflow:Froze\ 2\ variables.\\
			Converted\ 2\ variables\ to\ const\ ops.\\
			$
			
			Retraining on MobileNet:\\
				For MobileNet the accuracy was close to 82\%.\\
			$
			INFO:tensorflow:Final\ test\ accuracy = 81.6\%\ (N=98).\\
			INFO:tensorflow:Froze\ 2\ variables.\\
			Converted\ 2\ variables\ to\ const\ ops.\\
			$
			
			\textbf{Results: }
			We randomly picked 4 images from the testing set which our model had never seen before, 2 of which were of AD patients and 2 of the images were of control group. Both models were fed with one of each category of images for inference and for outputting their associated confidence scores as a result.
			
			Following figure shows the cross entropy, at beginning the cross entropy for MobileNet, was as high as 4 and for Inception, it was greater than 0.5. With the optimization performed at the end of each epoch consistently reduced the cross entropy over the training period; resulting in the cross entropy of less close to 0 for MobileNet and less than 0.1 from 6000th step of the training.
				
		\begin{figure}[!htbp]
				\centering
				\includegraphics[width=\textwidth,height=\textheight,keepaspectratio]{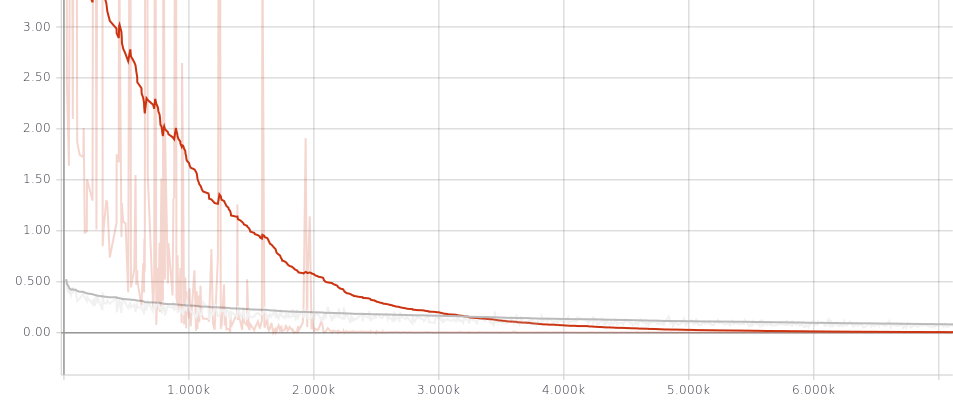}
				\caption{Cross entropy over the training period in epochs, Orange color representing MobileNet and Grey color representing Inception. The values are smoothened by the margin of 0.9, real values are shadowed in fainter color. X-axis is representing epochs in steps and Y-axis is representing cross entropy.}
		\end{figure}
	
			Following figure shows the accuracy, with the optimization performed at the end of each epoch, consistently increased the accuracy over the training period; resulting in the accuracy of more than 96\% for the retrained Inception model and the accuracy more than 99\% for the MobileNet model from 6000th step of the training.
			
		\begin{figure}[!htbp]
			\centering
			\includegraphics[width=\textwidth,height=\textheight,keepaspectratio]{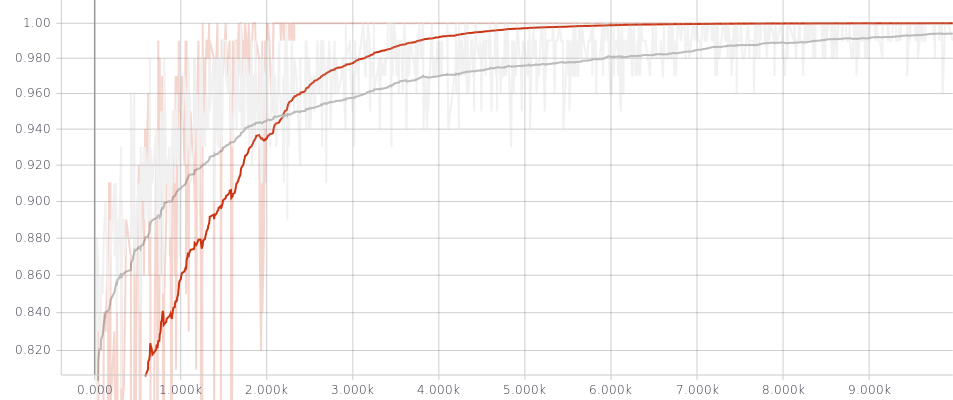}
			\caption{Shows accuracy over the training period in epochs, Orange color representing MobileNet and Grey color representing Inception. The values are smoothened by the margin of 0.9, real values are shadowed in fainter color. X-axis is representing epochs in steps and Y-axis is representing accuracy.}
		\end{figure}
	
		After achieving satisfactory accuracy, we froze both of the models as protobuff files with their parameters in order to preserve the learned features and graph structure, accompanied by their associated bottlenecks and training summaries. The models were then loaded for final execution/ inference purpose.
	
	Inference on Inception:\\
	For a random AD test sample, It gave out satisfactory scores.\\
	$
	Evaluation\ time\ (1-image): 4.007s\\
	alzheimer\ 0.724192\\
	control\ 0.275808\\
	$
	For a random Control test sample, It also gave out satisfactory scores.\\
	$
	Evaluation\ time\ (1-image): 4.009s\\
	control\ 0.999625\\
	alzheimer\ 0.000374888\\
	$
	
	Inference on MobileNet:\\
	For a random AD test sample, It gave out satisfactory scores.\\
	$
	Evaluation\ time\ (1-image): 1.453s\\
	alzheimer\ 1.0\\
	control\ 1.89051e-08\\
	$
	For a random Control test sample, It also gave out satisfactory scores.\\
	$
	Evaluation\ time\ (1-image): 1.484s\\
	control\ 1.0\\
	alzheimer\ 1.40792e-19\\
	$
	
	\subsection{BellCNN: }
	\textbf{Data preprocessing:}
	Data wrangling is the foundation of our efforts through which weve obtained a usable set of training and testing data. Weve created randomized arrays of the image data which are saved as numpy arrays with one-hot encoding in order to train the model to have a binary decision as an ouput.\\\\
	\textbf{Model architecture:}
	Our Model architecture of BellCNN takes inspiration from the shape of the Bell Curve in Gaussian distribution [14]. Our first CL starts with 32 filters, followed by a CL with 64 filters, followed by a CL with 128 filters, followed by a CL with 64 filters, and followed by the final CL with 32 filters; each CL is accompanied by a PL (Max), All CLs and PLs were used with RELU as a primary activation function. After the final CL-PL, We propagate the results through an FL of 1024 nodes with a dropout rate of 80\% (0.8) with SoftMax as the activation function in order to facilitate the binary classification done by the Final layer consisting of 2 nodes. And at the end, we iterated, optimized and trained our model with Adam optimizer [11]. \\\\
	$
	Model BellCNN():\\\\
	1.\ Prepare\ input\ layer\ as\ initial\ layer\\
	2.\ Set\ input\ of\ layer\ N,\ to\ output\ of\ layer\ N-1\\
	3.\ Set\ stride\ to\ 5\\
	4.\ Set\ activation\ to\ RELU\\
	5.\ Convolve\ with\ 32\ filters\\
	6.\ Max\ pool\\
	7.\ Convolve\ with\ 64\ filters\\
	8.\ Max\ pool\\
	9.\ Convolve\ with\ 128\ filters\\
	10.\ Max\ pool\\
	11.\ Convolve\ with\ 64\ filters\\
	12.\ Max\ pool\\
	13.\ Convolve\ with\ 32\ filters\\
	14.\ Max\ pool\\
	15.\ Flatten\ with\ Fully\ connected layer of 1024 nodes\\
	16.\ Set\ dropout\ rate\ to\ 0.8\\
	17.\ Set\ activation\ to\ SoftMax\\
	18.\ Output\ with\ final\ layer\ of\ 2\ nodes\\
	19.\ Set\ optimizer\ to\ Adam\\
	20.\ Regress\\\\\\\\
	FitModel (X, Y):\\\\
	1.\ Load\ model\ graph\ of\ BellCNN\\
	2.\ Reshape\ X,\ Set\ input\ X\\
	3.\ Set\ target\ Y\\
	4.\ Set\ epochs\ to\ N\\
	5.\ Take\ snapshots\ and\ write\ summaries\ with\ P\ intervals\\
	$
	
	Weve set the dropout rate to 0.8, meaning that in a randomized manner, every node will get activated only 80\% of the time and is turned off 20\% of the time.\\After performing dropout regularization on FL, we needed to optimize the error with a gradient descent optimizer. We chose adam optimizer instead of usual stochastic gradient descent algorithm, because of its adaptive nature. It computes adaptive learning rates for parameters from the n-2 and n-1 moments of the gradient resulting in better optimization when compared to the stochastic method [11].\\
	
	\begin{figure}[!htbp]
		\centering
		\includegraphics[width=\textwidth,height=\textheight,keepaspectratio]{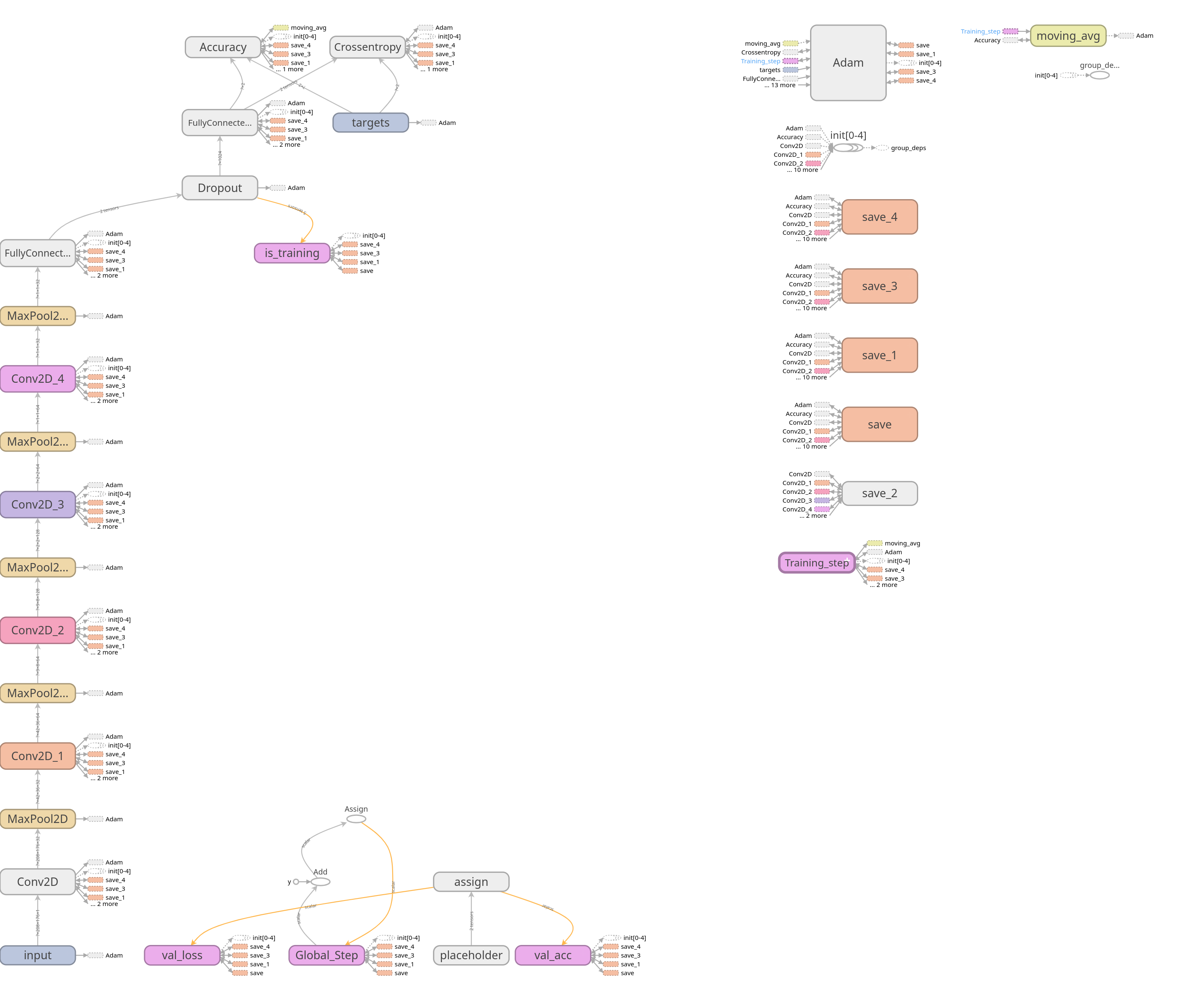}
		\caption{The Tensorboard visualization of  BellCNN as our frozen graph (model).}
	\end{figure}

	\textbf{Results: } We trained our model for more than 100 epochs with each epoch consisting of 7 batches and found that from 500th step our model was consistently giving good results.\\\\
	Following figure shows the loss, at beginning the loss was as high as 60\%, with the optimization performed at the end of each epoch consistently reduced the loss over the training period; resulting in the loss of less than 10\% from 500th step of the training.
	
	\begin{figure}[!htbp]
		\centering
		\includegraphics[width=\textwidth,height=\textheight,keepaspectratio]{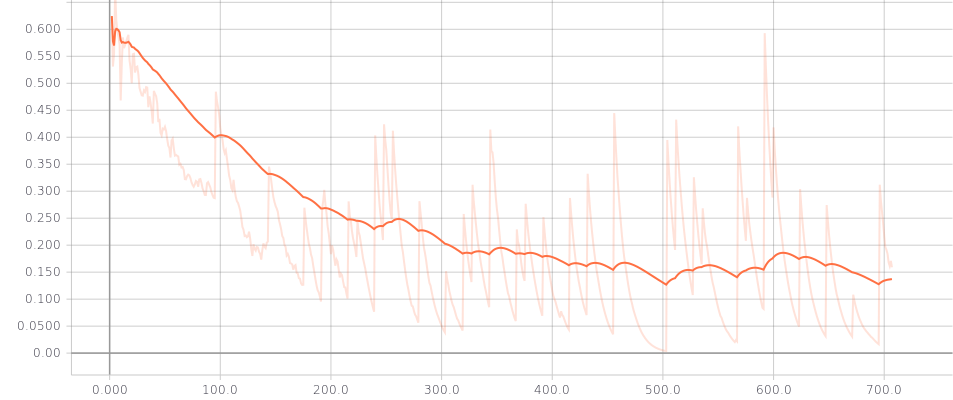}
		\caption{Shows loss over the training period in epochs, the values are smoothened by the margin of 0.9, real values are shadowed in fainter color. X-axis is representing epochs in steps and Y-axis is representing loss.}
	\end{figure}

	The loss was calculated by the categorical cross entropy method which is an efficient way of calculating loss in binary categories between predictions obtained and targets expected. This loss factor is taken into account by adam optimizer while updating weights and biased during the execution of back propagation. \\\\
	Following figure shows the accuracy, with the optimization performed at the end of each epoch, consistently increased the accuracy over the training period; resulting in the accuracy of more than 95\% from 500th step of the training.
	
		\begin{figure}[!htbp]
		\centering
		\includegraphics[width=\textwidth,height=\textheight,keepaspectratio]{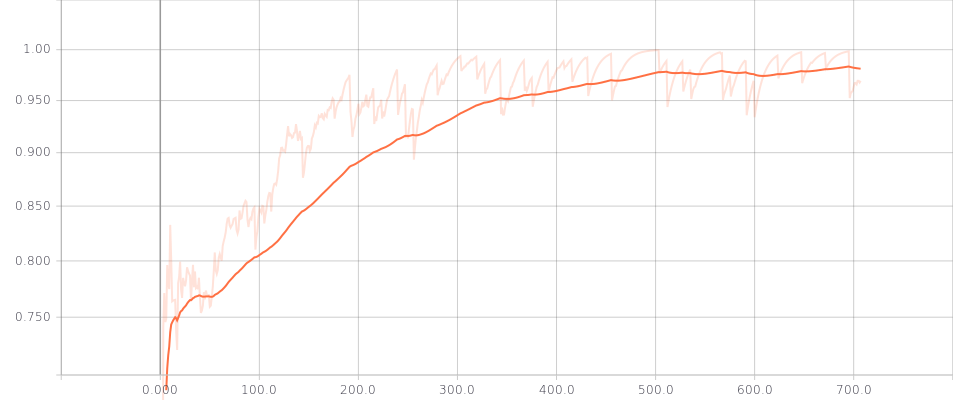}
		\caption{Shows accuracy over the training period in epochs, the values are smoothened by the margin of 0.9, real values are shadowed in fainter color. X-axis is representing epochs in steps and Y-axis is representing accuracy.}
		\end{figure}
	
	After achieving satisfactory accuracy, we froze the model as a protobuff file with its parameters in order to preserve the learned features and graph structure. The model was then loaded for final execution/ inference purpose.
	
	\begin{figure}[!htbp]
		\centering
		\includegraphics[width=\textwidth,height=\textheight,keepaspectratio]{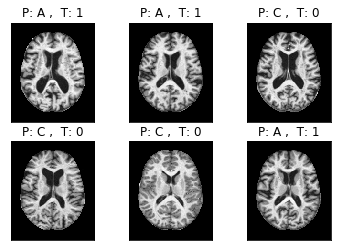}
		\caption{Shows classification results. Prediction is denoted by P; True labels are denoted by T. Prediction C denotes Control and Prediction A denotes Alzheimers Disease. True value 0 denotes Control and True value 1 denotes Alzheimers Disease.}
	\end{figure}

We had randomly picked 9 images from the testing set which our model had never seen before, 3 of which were of AD patients and rest of the images were of control group. Because of the dropout regularization and efficient data wrangling; we accurately predicted all of the test samples and obtained a more generalized model that can classify images of AD subjects with minimal error.

\section{DISCUSSION AND COMPARISON}
There are few difficulties faced by researchers such as: Resource heavy models, Expensive GPUs/ Cloud instances, Reliability of data. Among these difficulties, data is of utmost Importance; because bad data results into bad learning. Thus, correct and efficient data-wrangling is crucial. We used OASIS as our data source, so the reliability of our models are tied to the origin of our data source. 

Accuracy of the retrained Inception and MobileNet models is subject to scrutiny; as transfer learning should not be used in medical applications due to irregularities and a hidden potential of mishaps. Models obtained by transfer learning are also subject to influence by foreign input spaces. The better solution is to build the model from ground up dedicated to a single application. BellCNN does give acceptable results in terms of identification of AD and It is less likely to be influenced by a foreign input space.

\section{CONCLUSION}
We have shown that the proposed system has an ability to effectively pre-diagnose AD. And as this system is built on Tensorflow, we can scale our model to production instantly. Future work of this project aims to port and integrate the system to MRI devices and make the current system more usable. This goal, if achieved will also reduce the delay caused due to two independent systems having to manually inter-operate to get the same outcome.

\section{ACKNOWLEDGMENT}
We express our gratitude towards OASIS for making such an accurate and reliable dataset openly available for use. We express our gratitude towards Tensorflow team and the community supporting it; for making this project open source. Tensorflow is highly usable and has played a major role in democratizing data science and artificial intelligence. We express our gratitude towards continuum who made Anaconda available and helped data scientists get started with minimal efforts as most of the required utilities come pre-bundled. And we also express our gratitude towards known-unknown individuals and organizations that made this project possible.

\pagebreak
\noindent
\section*{REFERENCES}
Alzheimers Disease: https://www.nia.nih.gov/health/alzheimers-disease-fact-sheet\\\\
OASIS Dataset: http://www.oasis-brains.org/\\\\
Daniel S. Marcus, Tracy H. Wang, Jamie Parker, Open Access Series of Imaging Studies (OASIS): Cross-sectional MRI Data in Young, Middle Aged, Nondemented, and Demented Older Adults, Massachusetts Institute of Technology, Journal of Cognitive Neuroscience, 2007, 19:9, pp. 1498-1507\\\\
Daniel S. Marcus, Oasis fact sheet (rev. 2007-8-20) Cross-sectional data across the adult lifespan, 2007\\\\
Yann LeCunn, Leon Bottou, Yoshua Bengio, Patrick Haffner, Gradient-Based Learning Applied to Document Recognition, Proceedings of IEEE, November - 1998\\\\
Convolutional Neural Networks: https://www.tensorflow.org/tutorials/\\\\
Google Research, TensorFlow: Large-Scale Machine Learning on Heterogeneous Distributed Systems, arXiv:1603.04467v2 as https://arxiv.org/abs/1603.04467\\\\
A Guide to TF Layers: Building a Convolutional Neural Network, https://www.tensorflow.org/tutorials/layers\\\\
TF Transfer Learning: https://codelabs.developers.google.com/codelabs/tensorflow-for-poets\\\\
Alex Krizhevsky, Ilya Sutskever, Geoffrey E. Hinton, ImageNet Classification with Deep Convolutional Neural Networks,  University of Toronto, as Proceeding NIPS12 Proceedings of the 25th International Conference on Neural Information Processing Systems Pages 1097-1105\\\\
Diederik P. Kingma, Jimmy Ba, Adam: A Method for Stochastic Optimization, arXiv:1412.6980v9, https://arxiv.org/abs/1412.6980\\\\
Allan Raventos, Moosa Zaidi, Automating Neurological Disease Diagnosis Using Structural Mr Brain Scan Features, Stanford University\\\\
Saman Sarraf, Danielle D. DeSouza, John Anderson, Ghassem Tofighi, DeepAD: Alzheimers Disease Classification via Deep Convolutional Neural Networks using MRI and fMRI, Cold Spring Harbor Laboratory Press, DOI: https://doi.org/10.1101/070441\\\\
Gururaj Awate, Sunil Bangare, G Pradeepini, S Patil, Detection of Alzheimers Disease from MRI using Convolutional Neural Network with Tensorflow, arXiv:1806.10170v1, https://arxiv.org/abs/1806.10170
\end{document}